\documentclass[pra,twocolumn,superscriptaddress,10pt]{revtex4-1}
\usepackage{graphicx}
\usepackage{dcolumn}
\usepackage{bm}%
\usepackage[utf8]{inputenc}
\usepackage[mathlines]{lineno}
\usepackage{graphicx}
\usepackage[colorlinks=true,citecolor=red,urlcolor=black]{hyperref}
\usepackage{amsfonts}
\usepackage{amsmath,amssymb}
\usepackage{dsfont}
\usepackage{euscript}
\usepackage{float}
\usepackage{amsthm}
\usepackage{mathalfa}
\usepackage{bbm}
\usepackage{ragged2e}
\usepackage{color}
\usepackage{lipsum}
\usepackage{mathtools}
\usepackage{cuted}
\usepackage[dvipsnames]{xcolor}
\usepackage[normalem]{ulem}

\usepackage{xparse}
\usepackage{framed}
\usepackage{upquote}
\usepackage{tikz}
\usepackage{textcomp}
\DeclareMathSymbol{\mathdblquotechar}{\mathalpha}{letters}{`"}
\newcommand{\mathdblquote}{\mathtt{\mathdblquotechar}}
\begingroup\lccode`~=`"\lowercase{\endgroup
  \let~\mathdblquote
}
\AtBeginDocument{\mathcode`"="8000 }
\usetikzlibrary{matrix,backgrounds}
\usepackage{float}
\def\be{\begin{equation}}
\def\ee{\end{equation}}
\newcommand{\Gmat}{{G}}

\newcommand{\Tr}{\mathrm{ Tr }}

\newcommand{\sket}[1]{{\ensuremath{\lvert#1\rangle}}}
\newcommand{\lket}[1]{{\ensuremath{\left\lvert#1\right\rangle}}}
\newcommand{\ket}[1]{\if@display\lket{#1}\else\sket{#1}\fi}

\newcommand{\sbra}[1]{{\ensuremath{\langle#1\rvert}}}
\newcommand{\lbra}[1]{{\ensuremath{\left\langle#1\right\rvert}}}
\newcommand{\bra}[1]{\if@display\lbra{#1}\else\sbra{#1}\fi}

\newcommand{\sbraket}[2]{{\ensuremath{\langle#1\rvert#2\rangle}}}
\newcommand{\lbraket}[2]{{\ensuremath{\left\langle#1\!\left\rvert\vphantom{#1}#2\right.\!\right\rangle}}}
\newcommand{\braket}[2]{\if@display\lbraket{#1}{#2}\else\sbraket{#1}{#2}\fi}

\newcommand{\sketbra}[2]{{\ensuremath{\lvert #1\rangle\!\langle #2\rvert}}}
\newcommand{\lketbra}[2]{{\ensuremath{\left\lvert #1\right\rangle\!\!\left\langle #2\right\rvert}}}
\newcommand{\ketbra}[2]{\if@display\lketbra{#1}{#2}\else\sketbra{#1}{#2}\fi}

\newcommand{\proj}[1]{\ketbra{#1}{#1}}
\begin{document}
\preprint{APS/123-QED}
\title{Practical quantum key distribution with non-phase-randomized coherent states} 
\author{Li Liu}
\affiliation{State Key Laboratory of Networking and Switching Technology, Beijing University of Posts and Telecommunications, Beijing, 100876, China}
\affiliation{School of Science, Beijing University of Posts and Telecommunications, Beijing, 100876,  China}
\affiliation{Department of Electrical \& Computer Engineering, National University of Singapore, Singapore}
\author{Yukun Wang}%
\email{elewayu@nus.edu.sg}
\affiliation{Department of Electrical \& Computer Engineering, National University of Singapore, Singapore}
\author{Emilien Lavie}
\affiliation{Department of Electrical \& Computer Engineering, National University of Singapore, Singapore}
\author{Arno Ricou}%
\affiliation{Centre for Quantum Technologies, National University of Singapore, Singapore} 
\author{Chao Wang}
\affiliation{Department of Electrical \& Computer Engineering, National University of Singapore, Singapore}
\author{Fen Zhuo Guo}
\affiliation{State Key Laboratory of Networking and Switching Technology, Beijing University of Posts and Telecommunications, Beijing, 100876, China}
\affiliation{School of Science, Beijing University of Posts and Telecommunications, Beijing, 100876,  China}
\author{Charles Ci Wen Lim}\email{charles.lim@nus.edu.sg}
\affiliation{Department of Electrical \& Computer Engineering, National University of Singapore, Singapore}
\affiliation{Centre for Quantum Technologies, National University of Singapore, Singapore} 

\begin{abstract}
Quantum key distribution (QKD) based on coherent states is well known for its implementation simplicity, but it suffers from loss-dependent attacks based on optimal unambiguous state discrimination. Crucially, previous research has suggested that coherent-state QKD is limited to short distances, typically below 100 km assuming standard optical fiber loss and system parameters. In this work, we propose a six-coherent-state phase-encoding QKD protocol that is able to tolerate the total loss of up to 38 dB assuming realistic system parameters, and up to 56 dB loss assuming zero noise. The security of the protocol is calculated using a recently developed security proof technique based on semi-definite programming, which assumes only the inner-product information of the encoded coherent states, the expected statistics, and that the measurement is basis-independent. Our results thus suggest that coherent-state QKD could be a promising candidate for high-speed provably-secure QKD.
\end{abstract}

\maketitle

\section{Introduction}
Quantum key distribution (QKD)~\cite{BB84} is one of the most established quantum information technologies to date. Its basic goal is to distribute secret keys between two remote users (called Alice and Bob) embedded in an untrusted network. Importantly, unlike conventional key distribution methods, QKD is provably-secure and can be safely used with any cryptographic protocol that requires long-term security assurance. For an overview of QKD and its recent developments, we refer the interested reader to Refs.~\cite{Gisin2001,scarani2009,Lo2014}.

In prepare-and-measure QKD, there are generally two classes of coherent-state protocols, namely those that are based on phase randomization and those that give the phase reference information to Eve (the adversary). In the former, one first uniformly randomizes the phase of the coherent state $\ket{\sqrt{\mu}e^{i\theta}}$ to create a mixture of photon number states, i.e., 
\be \label{mixedphotons}
\rho_{\mu}=\int_0^{2\pi}\frac{d\theta}{2\pi} \proj{\sqrt{\mu}e^{i\theta}} =e^{-\mu}\sum_{n\geq 0} \frac{\mu^n}{n!}\proj{n}.
\ee
This mixture follows a Poisson distribution and emits a single-photon state with probability $\mu e^{-\mu}$ and multi-photon states with probability $1-e^{-\mu}(1+\mu)$, where $\mu$ is the mean photon number of the signal. Then, by using a statistical technique called the decoy-state method to bound the fraction of detected single-photon states, one can get secret key rates that are comparable to that with a true single-photon source~\cite{Hwang2003,lo2005decoy,XBWPRL2005}. In the following, we shall refer to phase-randomized coherent-state QKD as decoy-state QKD since they often mean the same protocol in practice.  

In the case that the phase reference is given to Eve, she no longer sees a mixture of photon number states, but a coherent state $\ket{\alpha_x}$ that is randomly drawn from a set of possible preparations $\{\ket{\alpha_x}\}_x$, which is necessarily linearly independent. Crucially, based on this phase information, Eve can optimize her attack strategy to distinguish between different subsets of the preparation set~\cite{Dsek2000,sun2012,tang2013}. In the worst case scenario (given the channel loss  is sufficiently high), she can determine the exact value of $x$ by performing unambiguous state discrimination. For this reason, most existing security analyses of non-phase-randomized coherent-state QKD show that the tolerable channel loss  is significantly below that of decoy-state QKD~\cite{yuen1996,Lo2007}. For example, assuming standard experimental settings, we find that the maximum distance (fiber length) of the phase-encoding coherent-state QKD protocol is generally shorter than 100 km, while decoy-state QKD can achieve up to about 250 km (see the simulation results below). 

However, in practice, it may be more attractive to consider non-phase-randomized coherent-state QKD. The main reason is that the set of security assumptions for non-phase-randomized coherent-state QKD is typically less stringent than that of decoy-state QKD. To appreciate this point, we note that it is very important for decoy-state QKD systems to completely randomize the phase of their quantum signals. If this assumption is not met, then there could be serious security loopholes~\cite{sun2012,tang2013}. Recently, it has been pointed out that the requirement of ideal continuous phase randomization can be replaced by discrete phase randomization~\cite{caoma2005}. This work represents an important step towards making decoy-state QKD more practical, however it introduces additional complexity into the security analysis, which can be tricky when considering finite-length keys. Thus, for practical reasons, it may still be of interest to investigate new coherent-state QKD protocols that can distribute secret keys over longer distances. 

In order to achieve good secret key rates using non-phase-randomized coherent states, it is important to have a tight estimation of Eve's information about the secret key. While in principle one could obtain a tight estimation of Eve's information by considering all the possible eavesdropping strategies, it is computationally intractable to fully characterize them, especially if the underlying Hilbert space dimension is unknown. To resolve this technical difficulty, here we employ the numerical tool introduced in Ref.~\cite{wang2019} to bound Eve's information, which roughly speaking, converts the characterization problem of Eve's strategies into a tractable hierarchy of semi-definite programs (SDP). It has been shown in Ref.~\cite{wang2019} that the SDP method provides a tighter bound on Eve's information as compared to existing methods that use the so-called quantum coin method~\cite{Lo2007}. Additionally, the SDP method is semi-device-independent (SDI)~\cite{Pawlowski2011,bowles2014,Ykwang2015,woodhead2015,himbeeck2017,hugo2019}, namely the detailed characterization of Bob's measurements is no longer necessary, but different to previous SDI which based on dimension or energy assumption,  the analysis here is made based on the known inner product information of the coherent states. 
As such, the security certified by this method is robust against a large class of quantum device flaws and imperfections.

In this work, we propose a six-coherent-state protocol and show that it offers significant advantages in both secret key rate and transmission distance over existing coherent-state QKD protocols~\cite{Lo2007,wang2019}. 
The organization of the paper is as follows: in Sec. \ref{section:Lo-Preskill protocol}, for pedagogical reasons, we first review a standard coherent-state QKD protocol that encodes the secret key into the relative phase of the coherent states. In Sec. \ref{section:Six states protocol}, we first introduce a six-coherent-state protocol which allows the test coherent states to use different mean photon number from that of the key states. Then, we use the method introduced in Ref.~\cite{wang2019} to derive the secret key rate of the protocol and simulate its expected performance using standard experimental parameters. Finally, we conclude our findings in Sec. \ref{section:Conclusion}.

 \section{Phase-encoding BB84 coherent-state QKD}
 \label{section:Lo-Preskill protocol}

 In this section, we first review a popular coherent-state QKD protocol, which is inspired by the celebrated Bennett \& Brassard 1984 (BB84) QKD~\cite{BB84}. Here, the protocol encodes the secret bit into the relative phase of two coherent states, where the first coherent state is the signal state (the modulated signal) and the second coherent state is the reference state (fixed and whose phase reference is given to Eve). More specifically, Alice sends one of four states randomly:
\begin{eqnarray}
&&\ket{\tilde{0}_{\rm{key}}}=\ket{\alpha}_R\otimes\ket{\alpha}_S,\nonumber\\
&&\ket{\tilde{1}_{\rm{key}}}=\ket{\alpha}_R\otimes\ket{-\alpha}_S,\nonumber\\
&&\ket{\tilde{0}_{\rm{test}}}=\ket{\alpha}_R\otimes\ket{i\alpha}_S,\nonumber\\
&&\ket{\tilde{1}_{\rm{test}}}=\ket{\alpha}_R\otimes\ket{-i\alpha}_S,
\end{eqnarray}
where the phase of $\alpha$ is relative to a fixed classical phase reference frame that Eve can access. Here, the subscript $S$ and $R$ denote the signal state and reference state, respectively. Then, Bob measures the signal he receives in either key basis or test basis. When Bob measures the signal in key basis, he combines the two modes in an interferometer, and directs modes $a_{0,\rm{key}}$, $a_{1,\rm{key}}$ to two different threshold photon detectors, where
 \begin{eqnarray}
 &&a_{0,\rm{key}}=(a_R+a_S)/\sqrt2,\nonumber\\
 &&a_{1,\rm{key}}=(a_R-a_S)/\sqrt2.
\end{eqnarray}
In ideal case, if Alice sends $\ket{\tilde{0}_{\rm{key}}}$ to Bob and he measures in the key basis, detector $D_0$ (model $a_{0,\rm{key}}$) will click (with some probability depending on the channel loss ) and detector $D_1$ (model $a_{1,\rm{key}}$) will not click. Likewise, if Alice sends $\ket{\tilde{1}_{\rm{key}}}$, then detector $D_0$ will not click and detector $D_1$ will click with some probability. Hence, Bob can determine the bit value that Alice encodes in key basis. When Bob measures in the test basis, he directs modes $a_{0,\rm{test}}$, $a_{1,\rm{test}}$ to the detectors, where
 \begin{eqnarray}
 &&a_{0,\rm{test}}=(a_R+ia_S)/\sqrt2,\nonumber\\
 &&a_{1,\rm{test}}=(a_R-ia_S)/\sqrt2.
\end{eqnarray}
Similarly, Bob can determine the bit value that Alice encodes in test basis. Then, Alice and Bob retain only the successful events in which they choose the same basis through public communication. After performing error correction step and privacy amplification step, Alice and Bob obtain the final secure key.

To prove the security of the protocol, one can either use the quantum coin method~\cite{Lo2007} or the SDP method proposed in Ref.~\cite{wang2019}. It has been demonstrated in the latter that the SDP method provides a tighter security analysis. Interestingly, Ref.~\cite{wang2019} also suggested that it may be useful to vary the test coherent states. In particular, the authors demonstrated that, in the case of coherent-one-way QKD, it is possible to significantly extend the key distribution distance by varying the intensity of the test coherent state in the protocol. In light of this observation, we propose the following phase-encoding QKD protocol. 

\section{Six-coherent-state QKD}
\label{section:Six states protocol}
\begin{figure*}[ht]
\includegraphics[scale=0.25]{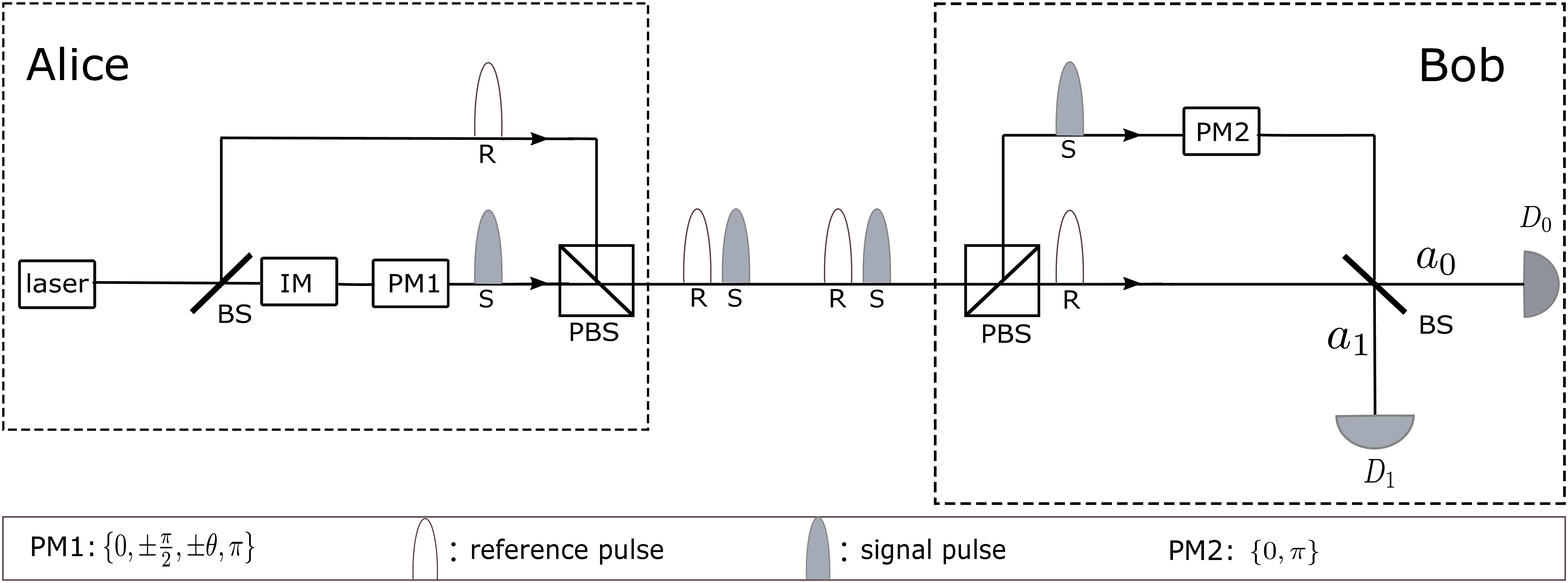}
\caption{\label{schematic} Schematic of the six-coherent-state QKD protocol.  The intensity modulator (IM) and phase modulator 1 (PM1) is positioned in Alice's site to prepare the intended coherent states, $\ket{\psi_z}$. Phase modulator 2 (PM2) in Bob's site is used to randomly rotate the phase of the signal signal by 0 or $\pi$. To avoid reducing the intensities of the coherent states, we can either use polarization multiplexing method as shown in the figure, or deploy active optical switches.
}
\end{figure*}

Here, we detail our proposed six-coherent-state protocol, and illustrate one of its possible implementation schemes in Fig.~\ref{schematic}.

\begin{figure}[ht]
\includegraphics[scale=0.65]{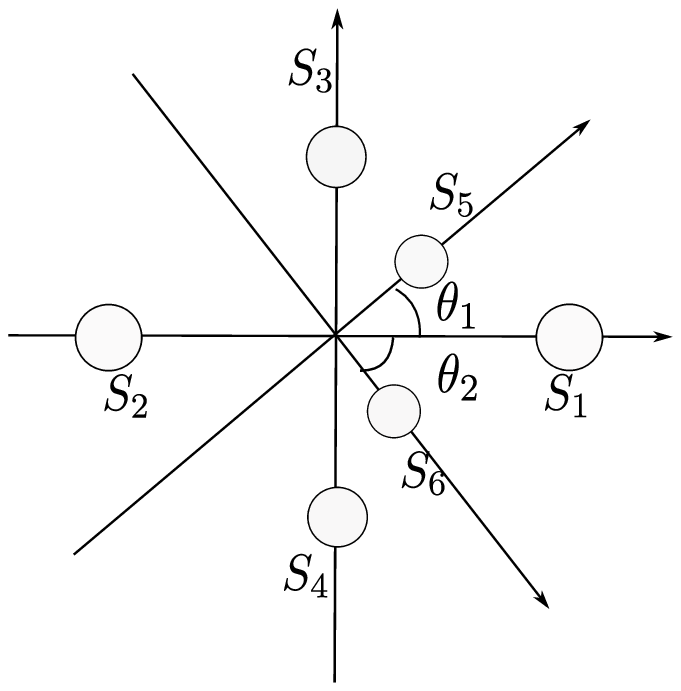}
\caption{\label{states} Distribution of coherent states in phase space. The key basis represents the real part of the space. Here, the intensities and phases $\{\theta_1,  \theta_2\}$ are given by the optimization: in fact, the optimization suggests that the optimal encoding scheme is given by $\mu_1=\mu_2=\mu_3$ and $\theta_1= \theta_2=\pi/2$.}
\end{figure}

\noindent\emph{Preparation}.-Alice randomly prepares one of those six quantum code states $\{\ket{\psi_z}\}_{z=1}^6$: $\ket{\psi_1}=\ket{\sqrt{\mu_1}}_R\ket{\sqrt{\mu_1}}_S$, $\ket{\psi_2}= \ket{\sqrt{\mu_1}}_R\ket{-\sqrt{\mu_1}}_S$, 
$\ket{\psi_3}=\ket{\sqrt{\mu_2}}_R\ket{i\sqrt{\mu_2}}_S$, $\ket{\psi_4}=\ket{\sqrt{\mu_2}}_R \ket{-i\sqrt{\mu_2}}_S$, 
$\ket{\psi_5}=\ket{\sqrt{\mu_3}}_R\ket{e^{i \theta_1}\sqrt{\mu_4}}_S$, $\ket{\psi_6}=\ket{\sqrt{\mu_3}}_R\ket{e^{-i \theta_2}\sqrt{\mu_4}}_S$, shown in Fig.~\ref{states}. Then, she sends it to Bob via the quantum channel. The states $\ket{\psi_1}$ and $\ket{\psi_2}$ are used to extract keys, while the rest are used to estimate Eve's information about the secret key. In this case, the key bit is encoded in the relative phase of the signal state and reference state. Here, the intensities and phases are free parameters, which will be used to optimize the secret key rate.

\noindent\emph{Measurement}.-Upon receiving the state, Bob directs it to the interferometer to recover the bit information. We use positive-operator valued measures (POVMs) to describe Bob's interference operations: Bob randomly performs one of two possible POVMs denoted by $\{B_y^b\}$,  where $y\in\{0,1\}$ represents basis choice. Taking signal loss and detection inefficiency into account, each measurement has three possible outcomes $b\in\{0,1,\emptyset\}$, where $\emptyset$ represents the empty detection event. In the event that both detectors click, Bob assigns a random bit to it. Additionally, we require that Bob's measurements satisfy basis-independent assumption: measurement operators corresponding to detection loss are the same for both measurement settings, i.e. ${B_0^\emptyset}={B_1^\emptyset}$. This is to ensure that the probability of detecting a signal is independent of Bob's measurement choice, which is necessary to rule out detection side-channel attacks exploiting the channel loss ~\cite{lydersen2010}. Thus Bob's three-outcome POVM is equivalent to a two-outcome POVM that determines the key bit, preceded by a basis-independent “filter”.

\noindent\emph{Parameter estimation and key distillation}.-Alice and Bob announce their basis choices through the public channel and decide if the error rates fulfill certain thresholds. If the test is successful, then Alice and Bob proceed with error correction and privacy amplification to extract a secret key. 

To compute the security of the protocol, we can use a certain entropic uncertainty relation for quantum memories~\cite{Berta2010} and Fano's inequality~\cite{Cover2006}. Using these, it can be shown that the asymptotic secret key rate of QKD against collective attacks is
\begin{eqnarray} \label{skr}
R^\infty_{\rm{key}}\ge \max \{0,p_{\rm{det}}(1-h_2(e_{\rm{bit}})-h_2(e_{\rm{ph}}))\},
 \end{eqnarray}
 where $e_{\rm{bit}}$ and $e_{\rm{ph}}$ are the bit error rate and phase error rate of the key basis, $p_{\rm{det}}$ is the probability of detection in key basis, and $h_2(\cdot)$ is the binary entropy function. The extension to general attacks is then achieved by using proof techniques like the post-selection technique~\cite{christandl209} or entropy accumulation theorem~\cite{dupuis2016}. These results imply that it is sufficient to consider security against collective attacks. 

\subsection{SDP optimization problem for bounding Eve's information}

From Eq. \eqref{skr}, we see that the secret key rate is obtained once we know the detection probability, bit error rate, and the phase error rate. To obtain these values, we first consider that Alice, Bob and Eve share a purified tripartite state $\ket{\Phi}$ after the transmission,  
\be
\ket{\Phi}=\frac{\ket{+}_A\ket{\phi_1}_{BE}+\ket{-}_A\ket{\phi_2}_{BE}}{\sqrt{2}}.
\ee
where Eve's set of possible operations has been considered in $\ket{\phi_1}_{BE}$ and $\ket{\phi_2}_{BE}$.
Conditioned on Bob observing a successful detection, the bit error rate of the key basis is then given by 
\begin{eqnarray} \nonumber
e_{\rm{bit}}&=&\frac{\bra{\Phi}(\ketbra{+}{+}\otimes B_0^1+\ketbra{-}{-}\otimes B_0^0)\ket{\Phi}}{p_{\rm{det}}}\\
&=&\frac{\bra{\phi_{1}}B_0^1\ket{\phi_{1}}+\bra{\phi_{2}}B_0^0\ket{\phi_{2}}}{2p_{\rm{det}}}.
\end{eqnarray}
The phase error rate in this case is the bit error rate that Alice and Bob would have observed if $\ket{\Phi}$ has been measured in the $\{\ketbra{+i}{+i},\ketbra{-i}{-i}\}$ basis by Alice and $B_1$ basis by Bob. Mathematically, this is given by
\begin{eqnarray} \nonumber
e_{\rm{ph}}&=&\frac{\bra{\Phi}(\ketbra{+i}{+i}\otimes B_1^0+\ketbra{-i}{-i}\otimes B_1^1)\ket{\Phi}}{2p_{det}}\\
&=& \frac{1}{2}-{\rm{Im}}\left\{\frac{\bra{\phi_{1}}B_1^0\ket{\phi_{2}}-\bra{\phi_{1}}B_1^1 \ket{\phi_{2}}}{2p_{\rm{det}}}\right\}.
\end{eqnarray}
One can see that $p_{\rm{det}}$ and $e_{\rm{bit}}$ are experimentally accessible, while the phase error rate $e_{\rm{ph}}$, which related to Eve's information about the secret bit is not. To get a tight estimation of $e_{\rm{ph}}$, in principle one should optimize over Eve's set of possible operations, namely captured in the transformed states $\ket{\phi_1}_{BE}$ and $\ket{\phi_2}_{BE}$ under the constraints imposed by the expected statistics. However, as we mentioned earlier, this type of characterization problem is often intractable or impossible to solve directly. 

Here, we employ the numerical tool introduced in Ref.~\cite{wang2019} to estimate the phase error rate, which converts the initial characterization problem into a hierarchy of semi-definite programs. More specifically, each hierarchy forms a convex set which outer-approximates the quantum set. Importantly, by going to higher hierarchies, the approximation becomes tighter. The quantum set here means the set of measurement statistics compatible with the set of prepared quantum signals and Bob's measurements. Thus, in using this method, we can optimize over the various convex sets to bound Eve's information, which in turn, provides a lower bound on the secret key rate of the protocol. 

With this SDP method, the detailed characterization of the quantum signals and measurements, including their dimension, is no longer required in the analysis.  Therefore, the transmission channel can be seen as an isometric evolution in higher dimension that takes the initial quantum signal state \ket{\psi_z} to some pure output signal state \ket{\phi_z}, which is shared between the receivers Bob and the network environment (possibly Eve). The inner product of the output states is preserved after the transmission, i.e. $\braket {\phi_z}{\phi_z'}=\braket {\psi_z}{\psi_z'}=\lambda_{zz'}$.
In the same way, the POVMs $\{B_y^b\}$ can be assumed as projective measurements in higher dimension. Then, we say that the probabilities of observing outcomes $b$ given setting $y$ and $ z$ admits a quantum system, if there exist a quantum state $\ket{\phi_z}$ and operators $B_y^b$ such that
\begin{eqnarray}
p(b|y,z)=\bra{\phi_z}B_y^b\ket{\phi_z}.
 \end{eqnarray}
where the  operators $\{B_y^b\}$ follow the  properties:
\begin{equation*}
\begin{split}
&(\emph{i}) \;\; \text{for}\; \text{any}\; y, B_y^b B_y^{b'}=0, \forall\; b\neq b',\\ 
&(\emph{ii})\; \Sigma_b{B_y^b}=\mathbb{I},\\ &(\emph{iii})\; (B_y^b)^2=B_y^b=(B_y^b)^{\dagger}.
\end{split}
\end{equation*}

\begin{figure}
\includegraphics[scale=0.16]{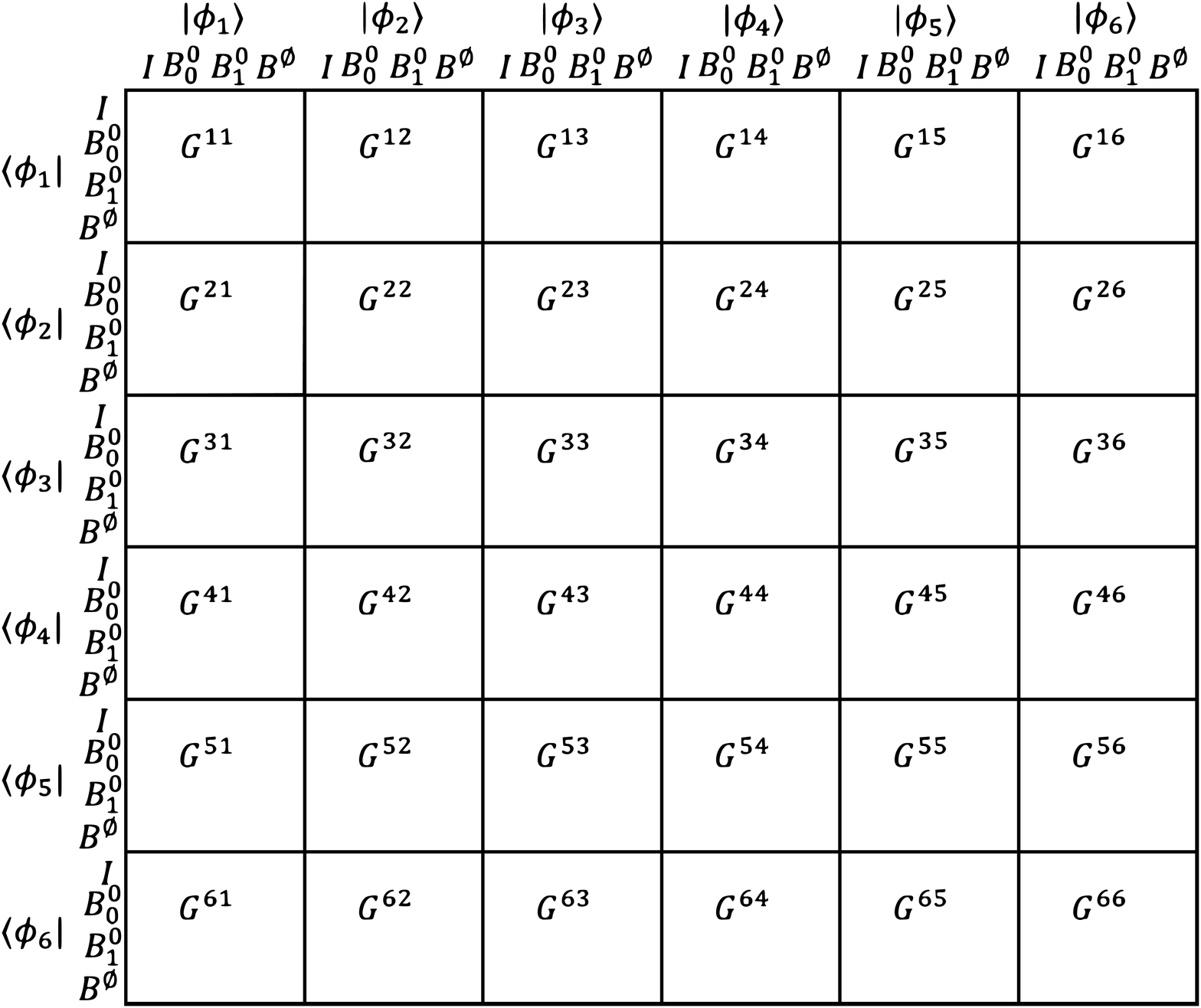}
\caption{\label{Gram} The Gram matrix under six quantum states and the operators set of $\mathcal{S}=\{\mathbb{I},B_0^0,B_1^0,B^{\emptyset}\}$. We have assumed that ${B_0^\emptyset}={B_1^\emptyset}={B^\emptyset}$. Actually this matrix is equivalent to the Gram matrix formed by the set $\{B_0^0,B_0^1,B_1^0,B_1^1,B^{\emptyset}\}$, since the correlations related to $B_y^1$ can be represented by $\mathbb{I}-B_y^0-B^{\emptyset}$. The reason we consider $\mathcal{S}$ is that it can reduce the matrix dimension.}
\end{figure}

A family of necessary conditions satisfied by the quantum observed probabilities thus can be introduced. 
Denote $\mathcal{S}=\{S_1,...,S_m\}$ as a finite set of $m$ operators, where each element is a linear combination of products of $\{B_y^b\}$. We define the $nm\times nm$ block Gram matrix $\Gmat$:
\begin{eqnarray}
\begin{aligned}
&\Gmat=\Sigma_{z,z'=1}^{n}G^{zz'}\otimes\ket{e_z}\bra{e_{z'}},\\
&\text{with,}\;\Gmat_{(i,j)}^{zz'}=\bra{\phi_z}S_i^{\dagger}\cdot S_j\ket{\phi_{z'}}\\
\end{aligned}
 \end{eqnarray}
 in which $\Gmat_{(i,j)}^{zz'}$ is defined as the inner-product of the vectors $\bra{\phi_z}S_i^{\dagger}$ and $S_j\ket{\phi_{z'}}$, for all $z,z'\in[n]$, $i,j\in[m]$. $\Gmat_{(i,j)}^{zz'}$ is the $ij$-entry of the matrix $\Gmat^{zz'}$ and $\{\ket{e_z}\}_{z=1}^n$ represents the standard orthonormal basis of $\mathbb{R}^n$. Evidently, the matrix $\Gmat$ is Hermitian and positive-semi-definite (PSD)~\cite{Horn2013} by definition. 
 
\begin{figure}[ht]
\includegraphics[scale=0.15]{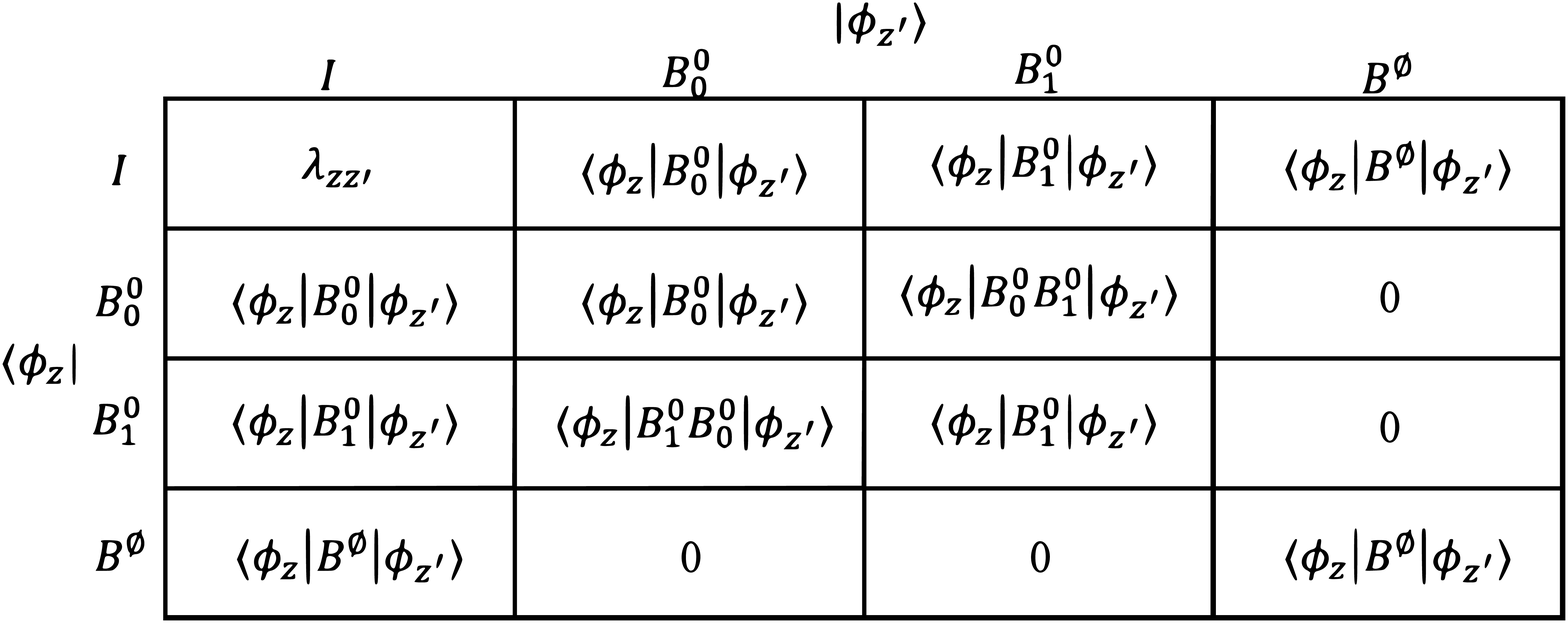}
\caption{\label{sub-Gram} The sub-block of the Gram matrix.}
\end{figure}

Taking $\mathcal{S}=\{\mathbb{I},B_0^0,B_1^0,B^{\emptyset}\}$ for example, the matrix $\Gmat$ is depicted in Fig.~\ref{Gram}. The whole matrix $\Gmat$ is partitioned  into $6\times6$ sub-blocks  $\{\Gmat^{zz'}\}_{zz'}$ by the classifier $z$, where each sub-block has size $4\times4$ and is illustrated in  Fig.~\ref{sub-Gram}.
The entries of every sub-block $\Gmat^{zz'}$  partly reflect the properties of $(\emph{i})$ - $(\emph{iii})$ satisfied by the operators and the overlaps of the signal states. For instance,  property ($\emph{ii}$) implies that we can introduce the identity operator $\mathbb{I}$ and  remove  one  of  the  operators;   property ($\emph{i}$) implies $\Gmat^{zz'}_{2,2}=\bra{\phi_{z}}B_0^0B_0^{0}\ket{\phi_{z'}}=\bra{\phi_{z}}B_0^{0}\ket{\phi_{z'}}$ and $\Gmat^{zz'}_{2,4}=\bra{\phi_{z}}B_0^0B^{\emptyset}\ket{\phi_{z'}}=0$; meanwhile the property
($\emph{iii}$) implies $\bra{\phi_{z}}B_y^bB_y^{b'}\ket{\phi_{z}}=(\bra{\phi_{z}}B_y^{b'}B_y^b\ket{\phi_{z}})^{\dagger}$.  In addition, since the overlaps of the code states are known, we have $\bra{\phi_{z}}\mathbb{I}\ket{\phi_{z'}}=\lambda_{zz'}$.

From the definition, each set $\mathcal{S}$ of operators 
yields a different Gram matrix $\Gmat$ and different linear constraints. The choice of
a particular $\mathcal{S}$ may seem arbitrary, but not all operators in $\mathcal{S}$ are independent. Moreover, they can be organized in a hierarchical
structure, such that  $\mathcal{S}$ can be  defined inductively,
\begin{eqnarray}
\begin{aligned}
    &\mathcal{S}_1=\{\mathbb{I},B_y^b\}\\
    &\mathcal{S}_2=\mathcal{S}_1 \bigcup \{B^b_yB^{b'}_{y'}\}\\
    &\mathcal{S}_3= \mathcal{S}_2 \bigcup \{B^b_yB^{b'}_{y'}B^b_y\}\\
     &\mathcal{S}_4=\dots\\
\end{aligned}
\end{eqnarray}
Thus, with each increase of the hierarchy, not only the dimension of the Gram matrix becomes larger, but also more new linear constraints are introduced. For example,
\[
\begin{split}
\text {for} \;\;\mathcal{S}_2\;\;:&\;\bra{\phi_{z}}B_0^0B_1^0\ket{\phi_{z'}}=\bra{\phi_{z}}B_0^0\cdot B_0^0B_1^{0}\ket{\phi_{z'}},\\
&\bra{\phi_{z}}B_0^0B_1^0\cdot B_0^0\ket{\phi_{z'}}=\bra{\phi_{z}}B_0^0\cdot B_1^0 B_0^0\ket{\phi_{z'}},\\
&\bra{\phi_{z}}B_0^0B_1^0\cdot B_1^0B_0^0\ket{\phi_{z'}}=\bra{\phi_{z}}B_0^0B_1^0B_0^0\ket{\phi_{z'}},\dots \\
\text {for} \;\;\mathcal{S}_3\;\;:&\;\bra{\phi_{z}}B_0^0B_1^0\cdot B_0^0B_1^0\ket{\phi_{z'}}\\
&=\bra{\phi_{z}}B_0^0\cdot B_1^0 B_0^0B_1^0\ket{\phi_{z'}}\\
&=\bra{\phi_{z}}B_0^0 B_1^0 B_0^0 \cdot B_1^0\ket{\phi_{z'}},\\
&\bra{\phi_{z}}B_0^0B_1^0\cdot B_0^0B_1^0B_0^0\ket{\phi_{z'}}\\
&=\bra{\phi_{z}}B_0^0 B_1^0 B_0^0 \cdot B_1^0B_0^0\ket{\phi_{z'}},\dots\\
\end{split}
\]
We denote the quantum set by $\mathsf{Q(\lambda)}$, which is formed by the probability distributions admitting a quantum system and the overlaps of the code states. Moreover, let the set of probability distributions defined by hierarchy of step $n$ be denoted as $\mathsf{Q(\lambda)_n}$. Then, we have that $\mathsf{Q(\lambda)} \subseteq \mathsf{Q(\lambda)_n}$. In fact, it has been demonstrated in Ref.~\cite{wang2019} that in the limit of $n$ (i.e., $n\rightarrow \infty$), $\mathsf{Q(\lambda)_n}$ converges to $\mathsf{Q(\lambda)}$. Ultimately, this means that the secret key rate obtained by the SDP can only be tighter when considering higher hierarchies.

 To compute Eq.~\eqref{skr}, we thus only need to maximize $e_{\rm{ph}}$ using SDP under the conditions that $p_{\rm{det}}$ and $e_{\rm{bit}}$ are fixed to some experimental model. In particular, the SDP problem for maximizing the phase error rate is:
\begin{equation}
\begin{split}
\text{maximize}:&\;e_{\rm{ph}} \\
\text{subject to}:&\;\braket{\phi_z}{\phi_{z'}}=\lambda_{zz'}, {\forall z,z'}\\
&\; \Gmat\geq 0,\; \\
&\;\Tr(F_k \Gmat)=p_k,\\
&\;\Tr(R_k \Gmat)=g_k,\\
&\;e_{\rm{bit}},p_{\rm{det}}\  \text{fixed to experimental model}\\
\end{split}
\end{equation}
where $F_k$'s and $R_k$'s are constant matrices. Note that $F_k$'s are used to pick up the terms in $\Gmat $ which are associated with the observed distributions $p(b|x,y)$; meanwhile $R_k$'s are used to pick up the terms in $\Gmat $ which are associated with the linear constraints induced by the quantum operators and states.

\subsection{Numerical simulation}
\begin{figure}[ht]
\includegraphics[scale=0.42]{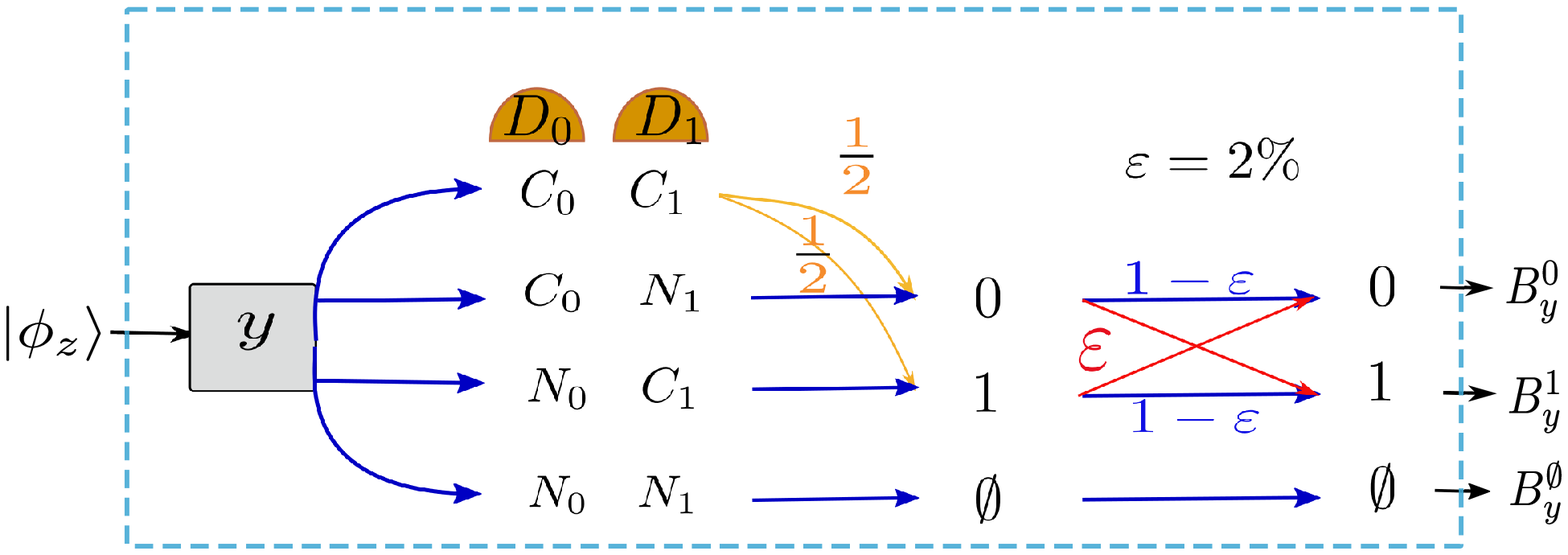} 
\caption{\label{caculation}Statistical model considering dark count and misalignment error. Where $y\in\{0,1\}$ represents the basis choice, $C_0$ and $C_1$ ($N_0$ and $N_1$) denote the event of detection $D_0$ and $D_1$ click (do not click), respectively. The notation  $"b"\in\{"0","1","\emptyset"\}$ represents the outcome without considering the misalignment error, while $b\in\{0,1,\emptyset\}$ denotes the outcome considering the misalignment error. }
\end{figure}

We simulate the secret key rate of the protocol using a realistic model, which includes total loss (including channel loss and detection efficiency), misalignment error of the optics, and the dark counts of single photon detectors. As Bob uses two threshold detectors, there are four possible outcomes when he measures a signal. By mapping double clicks randomly into 0 or 1, Bob's measurement realizes a POVM with three outcomes: 0, 1 and inconclusive. Note that the double clicks are interpreted as key bits~\cite{inamori2007,Lo2007}. In order to obtain a realistic detection model, we consider the error model illustrated in Fig.~\ref{caculation}.

We assume that the detector dark count rate is $p_{\rm{dc}}=10^{-7}$ and misalignment error rate is $\varepsilon=2\%$. For a given total loss $1-\eta$, the probability of detectors clicking and not clicking without misalignment error are
\begin{equation}
\begin{split}
N_j(z,y)=&(1-p_{\rm{dc}})[\frac{\bra{\sqrt{\eta}\phi_z}_R+(-1)^je^{-iy\frac{\pi}{2}}\bra{\sqrt{\eta}\phi_z}_S}{\sqrt{2}}\ket{0}\\
&\bra{0}\frac{\ket{\sqrt{\eta}\phi_z}_R+(-1)^je^{iy\frac{\pi}{2}}\ket{\sqrt{\eta}\phi_z}_S}{\sqrt{2}}],\\
C_j(z,y)=&1-N_j(z,y),
\end{split}
\end{equation}
where $j\in\{0,1\}$ represents detector $D_0$ and $D_1$, $y\in\{0,1\}$ is Bob's basis choice, $\ket{\phi_z}_S$ and $\ket{\phi_z}_R$ denote the signal and reference portion of the state $\ket{\phi_z} $ respectively.

Let $T("b",z,y)$ denotes the probability of getting the outcome $"b"$. We have that
\begin{equation}
\begin{split}
&T("0",z,y)=C_0(z,y)N_1(z,y)+\frac{C_0(z,y)C_1(z,y)}{2},\\
&T("1",z,y)=C_1(z,y)N_0(z,y)+\frac{C_0(z,y)C_1(z,y)}{2},\\
&T("\emptyset",z,y)=N_0(z,y)N_1(z,y).
\end{split}
\end{equation}
Considering the misalignment error $\varepsilon$, the probabilities of $B_y^b$ are

\begin{figure*}[t!]
\includegraphics[scale=0.8]{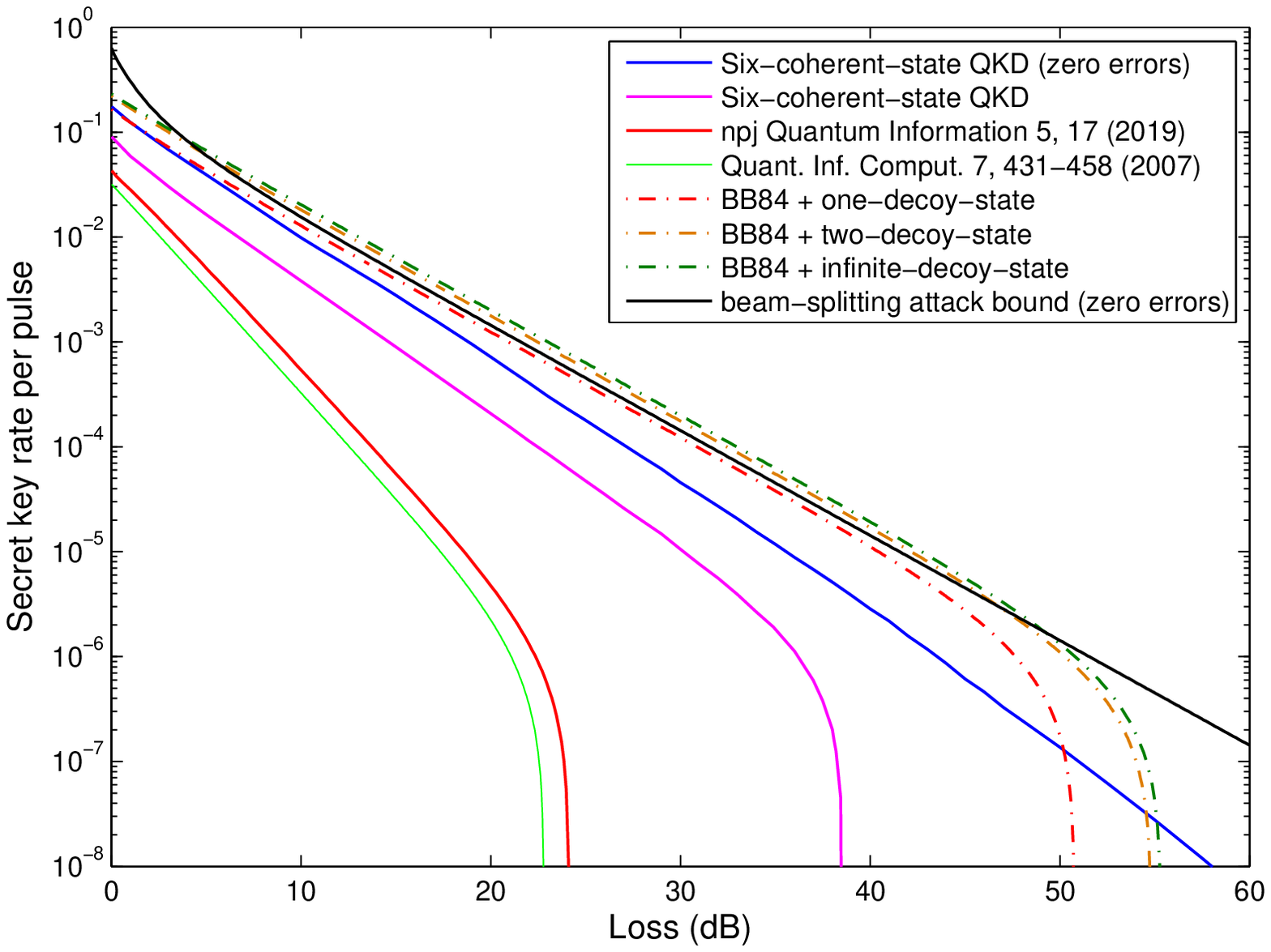}
\caption{\label{figcomparision} For the
key rate simulation, we assume a detector dark count rate of $p_{\rm{dc}}=10^{-7}$ and a misalignment error rate of $\varepsilon=2\%$. The comparison is based on the same error model and same experiment parameters. The protocol of BB84 plus decoy technology we considered here comes from Ref.~\cite{Alberto2018}. The one-decoy scenario of Ref.~\cite{Alberto2018} and our six-coherent-state protocol both have two key states, and four test states.  And we have less test states compared to more-decoy states BB84 protocol.  The optimization is carried out with the MATLAB packages YALMIP~\cite{lof2004} and the SDP solvers SEDUMI~\cite{sturm1999}.}
\end{figure*}
\begin{equation}
\begin{split}
&P(0,z,y)=(1-\varepsilon)T("0",z,y)+\varepsilon T("1",z,y),\\
&P(1,z,y)=(1-\varepsilon)T("1",z,y)+\varepsilon T("0",z,y),\\
&P(\emptyset,z,y)=T("\emptyset",z,y).
\end{split}
\end{equation}
\begin{table}
\caption{\label{probabilities} The statistics  of each POVM for each coming states. }
\begin{ruledtabular}
\begin{tabular}{cccccc}
\multicolumn{1}{c}{\textrm{}}&
\textrm{$B_0^0$}&\textrm{$B_0^1$}&\textrm{$B_y^{\emptyset}$}&\textrm{$B_1^0$}&\textrm{$B_1^1$}\\
\colrule
$\ket{\phi_1}$ & $P(0,1,0)$ & $P(1,1,0)$ & $P(\emptyset,1,y)$
& $P(0,1,1)$ & $P(1,1,1)$\\

$\ket{\phi_2}$ & $P(0,2,0)$ & $P(1,2,0)$ & $P(\emptyset,2,y)$
& $P(0,2,1)$ & $P(1,2,1)$\\

$\ket{\phi_3}$ & $P(0,3,0)$ & $P(1,3,0)$ & $P(\emptyset,3,y)$
& $P(0,3,1)$ & $P(1,3,1)$\\

$\ket{\phi_4}$ & $P(0,4,0)$ & $P(1,4,0)$ & $P(\emptyset,4,y)$
& $P(0,4,1)$ & $P(1,4,1)$\\

$\ket{\phi_5}$ & $P(0,5,0)$ & $P(1,5,0)$ & $P(\emptyset,5,y)$
& $P(0,5,1)$ & $P(1,5,1)$\\

$\ket{\phi_6}$ & $P(0,6,0)$ & $P(1,6,0)$ & $P(\emptyset,6,y)$
& $P(0,6,1)$ & $P(1,6,1)$\\
\end{tabular}
\end{ruledtabular}
\end{table}
Thus, the statistics of each POVM measurement on each coming states can be evaluated, as shown in Tab.~\ref{probabilities}. Accordingly, the probability of detecting a signal $p_{\rm{det}}$ for key basis and the corresponding bit error rate  $e_{\rm{bit}}$ are then given by
\begin{equation}
\begin{split}
p_{det}&=T("0",1,0)+T("1",1,0)\\
&=1-(1-p_{\rm{dc}})^2e^{-2\eta|\alpha|^2},\\
e_{bit}&=\frac{\varepsilon T("0",1,0)+(1-\varepsilon)T("1",1,0)}{p_{det}}\\
&=\frac{\frac{p_{dc}}{2}+\varepsilon(1-p_{dc})+(\frac{p_{dc}}{2}-\varepsilon)(1-p_{dc})e^{-2\eta|\alpha|^2}}{p_{det}}.
\end{split}
\end{equation}

Subsequently, we maximize $e_{\rm{ph}}$ over the set of compatible probabilities using the first level of the hierarchy and the results of the numerical optimization with errors (blue curve) and without
errors (rose curve) are shown in Fig.~\ref{figcomparision}. In the absence of errors, the tolerable total loss is extended to more than 55 dB, which is extremely close to the collective beam-splitting attack bound (an upper bound on the secret key rate). This suggests that the six-coherent-state QKD protocol is pretty robust against the total loss  in the absence of errors. Taking into account practical errors, the tolerable total loss reaches up to 38 dB, which is significantly higher than previous results~\cite{Lo2007,wang2019} (around 23, 25 dB respectively) assuming the same error model specified in Fig.~\ref{caculation}. As compared with the protocol in Ref.~\cite{wang2019}, additional two test states are included in our protocol to give better characterization of the quantum system between Alice and Bob. Thus, a tighter bound on Eve's information could be obtained.

Moreover, through the optimization, we observe that the secret key rate is optimized when $\mu_1=\mu_2=\mu_3$ and $\theta_1=\theta_2=\pi/2$. This is expected since  $\mu_1=\mu_2=\mu_3$ corresponds to larger overlaps between the six coherent states, which imply that it will be more challenging for Eve to distinguish them, meanwhile $\theta_1=\theta_2=\pi/2$  provides more characterization of the test basis ($B_1$), and thus, resulting in a higher secret key rate.  Therefore, in practice, one has to only modulate the amplitude between two levels ($\mu_1$ and $\mu_4$) which makes our protocol highly practical and suitable for high-speed implementation.

For completeness, we also plot the secret key rate, obtained using decoy-state method, of the protocol in Ref.~\cite{Alberto2018} that is based on time-bin encoding of phase-randomised coherent states. The asymptotic secret key rate is estimated using the formula given in Ref.~\cite{ma2005}, without taking into account of the statistical corrections of finite-length keys. Considering the same simulation parameters (i.e., total loss , misalignment error and dark count rate), the plot of secret key rate against loss of the decoy-state QKD protocol is shown in Fig.~\ref{figcomparision}. The comparison shows that secret key rate of our protocol is comparable to that of the one-decoy-state protocol in low loss regime. Specifically, the key rates differ by only one order of magnitude in the region when the loss is less than 25 dB. For further comparison, we also simulated the optimized secret key rates of BB84 protocol with 2 decoy states and infinite number of decoy states, and our protocol still shows appreciable performance in the short distance regime.

\section{Conclusion}
\label{section:Conclusion}

In conclusion, on the one hand, phase randomization is a critical assumption made in the analysis of most coherent state QKD protocols and any deviation from preparing perfectly phase-randomized coherent states may pose serious threats of a security breach. On the other hand, existing analysis on non-phase-randomized coherent states are overly pessimistic and yield secret key rates that are inferior to alternative protocols based on the decoy-state method. In this paper, we presented and analyzed the security of a six-coherent-state phase encoding QKD protocol based on non-phase-randomized coherent states. Our analysis requires fewer assumptions on both the quantum state preparation and measurement processes as it relies solely on the overlaps of the code states as well as the observed statistics. Simulating with realistic experimental model (dark count rate of $10^{-7}$ and misalignment error of $ 2\%$), our protocol can tolerate a total loss of up to 38 dB, which is much higher than previous works. Moreover, we observed that the secure key rates of our protocol are comparable with the BB84 decoy-state protocol in the low loss regime. In addition to the improved key rates, our protocol only requires the modulation of six relative phases and two amplitudes, which is easy to implement. Hence, we have shown that one could implement coherent state QKD without performing phase randomization and yield comparable key rates with other known protocols.

\begin{acknowledgments}
We would like to thank Hugo Zbinden and Goh Koon Tong for the  valuable comments and careful reading of the manuscript. This research is supported by the National Research Foundation (Singapore), the Ministry of Education (Singapore), the National University of Singapore,
 the Asian Office of Aerospace Research and Development, and the National Natural Science Foundation of China (Grant No.61572081 and No.61672110).
 Li Liu kindly acknowledges support from the China Scholarship Council.
 
\end{acknowledgments}

\appendix

\nocite{*}

\bibliography{apssamp}

\end{document}